# Phase Errors in mid-IR Arrayed Waveguide Gratings based on graded index SiGe/Si technology

**AINUR KOSHKINBAYEVA,[1,*] RÉGIS OROBTCHOUK,[2] AND PIERRE LABEYE [1]**

[1]*University Grenoble-Alpes, CEA-LETI, MINATEC, Grenoble F-38054, France*
[2]*Institut des Nanotechnologies de Lyon, Université de Lyon, INSA-Lyon, Villeurbanne F-69621, France*
*\*Corresponding author: ainurkosh@gmail.com*

**In this paper, we present a method for analysis of phase errors and power truncation in SiGe/Si graded index arrayed waveguide gratings (AWGs) operating in mid-infrared spectral range. Semi-analytical model used herein is based on Gaussian approximation of the modal field and Fourier Optics. This method is applied to study the correlation between crosstalk level, effective index deviation in array waveguides and power truncation in 3.4 µm, 4.5 µm, 5.7 µm and 7.6 µm central wavelength AWGs. We show that the impact of effective index variation is more critical for AWGs with smaller operational wavelengths and power truncation should not exceed 5% for these types of devices. In conclusion, we refer to experimental data of 4.5 µm, 5.7 µm and 7.6 µm central wavelength AWGs where a cross talk level of 24 dB is obtained, corresponding to relative errors of effective indices and equivalent path length deviations respectively lower than 2 $10^{-5}$ and  of 0.2 µm in 10 mm length of the devices.**

## 1. INTRODUCTION

Today's growth of integrated optics technology sets an increasing demand for broader transmission at lower crosstalk. The tendency towards miniaturization of integrated optics components for higher integration density induces such drawbacks as increasing impact of scattering losses at the waveguide boundaries and phase errors mostly due to sidewall roughness [1]. The design of integrated optics element is chosen depending on desired performance specifications of the application. In particular, preferences in spectral response of the arrayed waveguide grating (AWG) differ in mid-IR multi-gas sensing and telecommunication. The operation of AWG multiplexer is based on focusing of input wavelengths at the output channel by means of constructive interference [1]. The latter is achieved by constant light path difference in the array of waveguides of AWG designed for a specific spectral range.

The actual performance of the device differs from simulation mainly in higher crosstalk that arises due to several factors such as slight variations in width and height of waveguides as well as reduction of number of array waveguides of the device. Extensive analyses of phase error were presented for telecommunication spectral range. In particular, several studies covered lithography mask grid effect on phase errors [2], [3]. Lee et al. demonstrated the effect of power truncation as well as standard deviation of path length on crosstalk degradation for AWGs operating at telecom wavelength [2]. Mostly letters presented phase error analysis and experimental results of silica AWGs [4]–[7]. Statistical analysis of phase errors with maximum acceptable standard phase error deviation were studied as well [8], [9].

Although AWG operation principle is the same in various applications, the requirements to spectral response differ which leads to slightly differing approach in design. In telecom application, the goal is to achieve separate channel outputs with the lowest inter-channel crosstalk possible [1]; whereas in broad-band source for gas detection, the preference leans towards stronger overlap between channels for the uniform coverage of mid-IR target spectral range. CEA-Leti previously presented waveguides made of SiGe compound with a unique graded index triangular profile [14] that enable design and fabrication of low-loss integrated optics devices operating in a wide range of mid-infrared. In case of graded-index waveguides, another potential source of phase errors could arise due to the limit of estimation accuracy of effective index taken for AWG design.

Here we present the results of study of crosstalk degradation due to effective index variation and power truncation in mid-IR AWGs. To the best of our knowledge, it is the first report on phase error analysis for graded-index AWGs in the given spectral range. The phase noise simulation is done by a semi-analytical design tool tested on AWGs, presented earlier by CEA-Leti [10]–[13]. We demonstrate the correlation between effective index variation and crosstalk for 3.4 µm, 4.5 µm, 5.7 µm and 7.6 µm AWGs. We also show the relation between crosstalk level of these devices and power truncation, arising due to reduction of the number of waveguides in the array. Results of quantitative phase error analysis are presented for four AWGs for standard deviation of effective indices up to $7 \cdot 10^{-5}$ and power truncation up to 5%. Finally, we evaluate effective index variation of 5.7 μm AWG based on experimental data of Fourier Transform Infra-Red (FTIR) spectrometer measurements and deviations in the effective indices of 4.5 μm and 7.6 μm AWGs based on crosstalk levels reported previously [10], [12].

## 2. NUMERICAL MODELING

The schematic of AWG is illustrated in Fig. 1. It is built of input/output slabs, input/output channels and an array of single-mode waveguides. The operation principle of AWG has been extensively discussed earlier in [1], [10], so we focus on the modeling of phase errors.

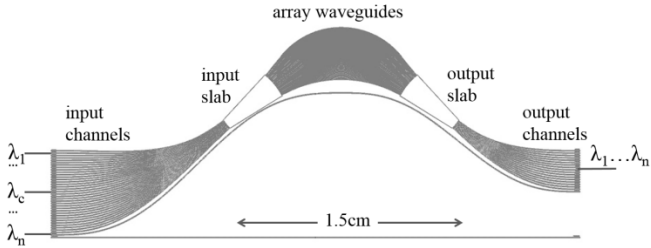

Fig. 1. Schema of AWG multiplexer.

Array waveguides are designed to support fundamental mode that is accurately approximated by Gaussian Fit. Modeling approach discussed in this paper greatly benefits from the simplification based on single-mode waveguides, that allow analytical calculation of mode coupling using overlap integral of Gaussian functions.

### A. Phase error modeling

As shown in Fig. 2, each subsequent array waveguide has a constant length difference ΔL, so each beam mode, propagating through waveguide array, acquires a phase shift. Input beam with certain wavelength ($\lambda_1,...,\lambda_n$) accumulates its corresponding phase difference Δφ in the array waveguides that causes a tilt of the phase front and defines a position of diffracted beam on the focal curvature Δ$s_{out}$ according to equation (1):

$$\Delta \varphi = \beta_{eff}(\sigma) \cdot \Delta L = 2\pi m - d \cdot \beta_s(\sigma) \cdot (\sin \theta_{in} + \sin \theta_{out}) \quad (1)$$

where $d$ is the waveguides spacing, $\beta_{eff}(\beta_s) = \frac{2\pi n_{eff}(n_s)}{\lambda}$ is the rib (slab) propagation constant, $n_{eff}(n_s)$ is the effective index of the array (slab) waveguide, $\lambda$ is the free space wavelength, $m$ is the diffraction order and $\theta_{in}$ ($\theta_{out}$) is the dispersion angle.

### B. Estimation of effective index variation

Sidewall roughness arising from fabrication imperfection alters effective index of the fundamental mode, which in turn causes phase errors [9]. To quantitatively evaluate this effect, we assumed phase errors to be random variables of normal distribution with zero mean value standard deviation. The phase component of the field in the array waveguides is then expressed as a sum $\sum_{i=1}^{N} \exp(-j(\Delta\varphi_i + \delta_i))$, where $i$ indicates the order of array waveguide, $N$ is the number of waveguides in the array, $\Delta\varphi_i$ is the phase shift of the ith waveguide in the array and $\delta_i$ is its corresponding discrepancy in phase conformable to $\sigma_\delta$ standard deviation of phase errors.

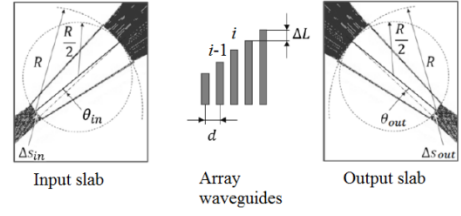

Fig. 2. Schemes of AWG input and output slabs (leftmost and rightmost, respectively), and array waveguides (center).

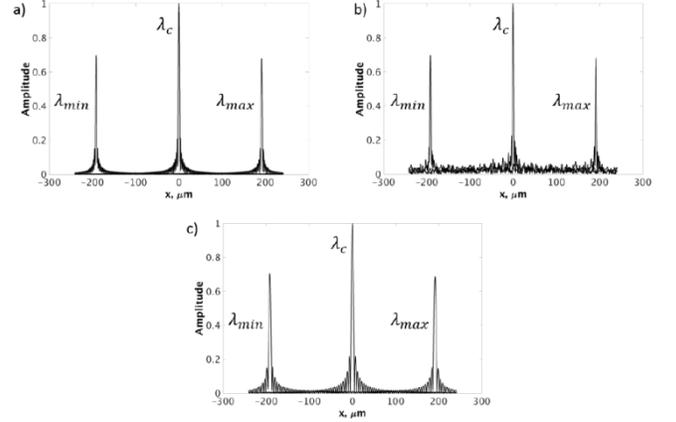

Fig. 3. Field amplitudes of 5.7 μm AWG at the focal line of the output slab before coupling to output waveguides: a) 0% truncation, zero level of phase errors; b) 0% truncation, standard deviation of phase errors $\sigma\varphi$=0.0009 rad; and c) 5% truncation, zero level of phase errors.

The introduction of phase errors to the model leads to increase of crosstalk level of AWG spectral response. For the given standard deviation of phase errors $\sigma_\delta$, one can evaluate effective index variation $\sigma_{\Delta neff}$ as follows:

$$\sigma_{\Delta neff} = (\lambda \sigma_\delta)/2\pi L \quad (2)$$

where $L$ is the path length segment under consideration.

### C. Modeling power truncation

As it is seen, phase errors mainly arise due to effective index variation of array waveguides. Reducing the number of waveguides could be an attractive option for achieving both the decrease of phase errors as well as scaling down the overall multiplexer dimensions. However, there is a trade-off between diminishing the number of array waveguides and retaining the sufficient amount of input power coupled from the slab. So, along with phase errors, we studied the effect of power truncation that appears when the number of array waveguides is reduced below the optimum defined by the width of the diverged beam in the input slab $w_{sx}$. The latter is approximated by Kirchhoff diffraction integral. Then, the power in the array waveguides can be evaluated as follows:

$$P = P_0 \cdot (1 - \exp(-2r^2/w_{sx}^2)), \quad (3)$$

where $P_0$ is the input power, $P$ is the power coupled to array waveguides and $r$ is the length of the focal line formed by array grating. The results of calculations are given in the following section.

## 3. RESULTS AND DISCUSSION

Using our home-made semi-analytical tool based on Fourier Optics we studied correlation between level of crosstalk, phase errors and power truncation in order to determine the fabrication tolerance required to obtain desired performance of the AWG.

Distortion of field amplitude under the impact of phase errors and power truncation first can be seen on the focal line, at the interface of output slab and output channels, right after the diffraction of beams. Fig. 3 illustrates an example for the case of minimum $\lambda_{min}$, central $\lambda_c$ and maximum wavelengths $\lambda_{max}$ of 5.7 μm AWG that can be generalized for other wavelengths and AWGs discussed in this paper.

The cumulative effect of phase errors (with standard deviations of 0 rad, 0.0002 rad, 0.0009 rad, 0.0017 rad, 0.0052 rad and 0.0087 rad) and power truncation (<0.1%, 1% and 5%) on performance of 3.4 μm, 4.5 μm, 5.7 μm and 7.6 μm AWGs were also analyzed. The phase error deviations were recalculated into deviations of effective indices of waveguides for $L$=10 mm length according to equation (3). In Fig. 4, graphs illustrate effective index error vs. crosstalk distribution at certain level of power truncation. Each data point represents an average of five independent calculations. Parameters of AWGs under consideration are given in Table 1.

It should be noted that the effect of power truncation is noticeably weighty. Given zero phase error level, all four AWGs exhibit minimum -70 dB crosstalk at optimum number of array waveguides, i.e. 185/205/258/259 array waveguides in 3.4 μm/4.5 μm/5.7 μm/7.6 μm AWG, respectively. As the power truncation level reaches 1%, i.e. 111/125/157/157 array waveguides in 3.4 μm/4.5 μm/5.7 μm/7.6 μm AWGs, respectively, the crosstalk levels become -34 dB for 3.4 μm and -38 dB for the rest AWGs. At 5% power truncation level, i.e. 90/101/125/127 array waveguides in 3.4 μm/4.5 μm/5.7 μm/7.6 μm AWG, respectively, the crosstalk is around -27 dB. The proximity of number of array waveguides in case of 5.7 μm and 7.6 μm AWGs could be understood by the choice of wider waveguide width for the latter. According to expectations, as the power truncation increases, the crosstalk becomes less sensitive to effective index variation. For the power truncation less than 0.1%, the impact of phase errors is the strongest, and the crosstalk degradation is as high as -30 dB and more for $5 \cdot 10^{-6}$ effective index variation at all four wavelengths. It should be noted, that for infrared gas sensing application, the crosstalk should preferably be below -20 dB. As shown in Fig. 4, given the crosstalk requirement, there is no need to investigate power truncation over 5%, as the crosstalk as of now approaches the acceptable limit. It is clearly seen, that the impact of effective index variation is more critical for AWGs with smaller operational wavelengths.

By reverse calculation, knowing the crosstalk level of a device, one can estimate the effective index deviation. Here we refer to 4.5 μm [10] and 7.6 μm [12] AWGs reported previously, where the crosstalk levels were around -20 dB with power truncation of about 1%. From equation (2), relative error of effective index $\sigma_{\Delta neff}/n_{eff}$ for 10 mm waveguide is $\approx 1.033 \times 10^{-5}$ for 4.5 μm AWG and $\approx 1.764 \times 10^{-5}$ for 7.6 μm AWG. The phase error deviations could also be represented in equivalent path length deviations, that is approximately 0.10 μm in 4.5 μm AWG and 0.20 μm in 7.6 μm AWG length (or $\approx \lambda/40$), which is noticeably smaller compared to 10 mm length of the device. Fig. 5 presents experimental data of AWG with central wavelength at 5.7 μm. Measured spectrum is shown with blue solid line and simulation – with green. The crosstalk of the device is 24 dB, which corresponds to relative error of effective index of $1.027 \times 10^{-5}$ or $\approx 0.10$ μm.

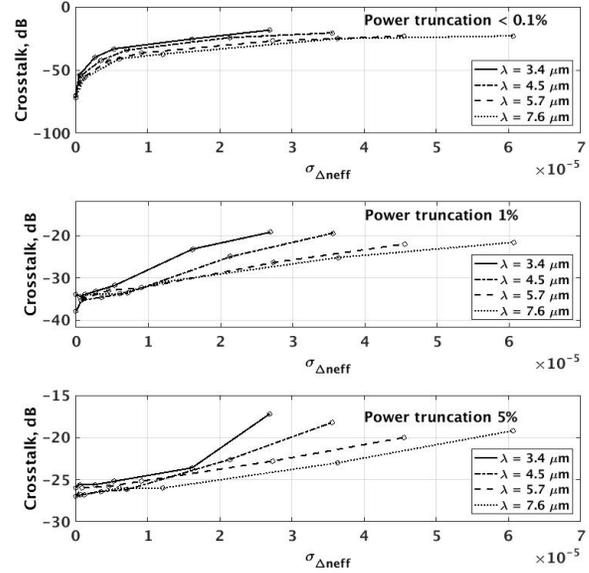

Fig. 4. Impact of standard deviation of effective index (10 mm waveguide) and power truncation on crosstalk.

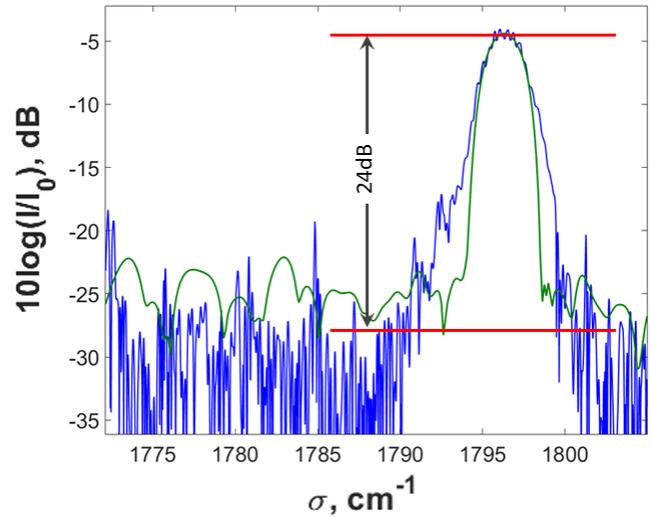

Fig. 5. 27th channel output of 5.7 μm AWG: blue line – measured spectrum, green line – simulation. Crosstalk level is 24 dB.

| $\lambda_c$, (μm) | $\Delta\lambda$ (nm) | $N_{ch}$ | $D_{AWG}$, (μm/cm-1) | w / h, (μm) | $n_g(\lambda_c)$ |
|---|---|---|---|---|---|
| 3.4 | 182 | 9x19 | 0.79 | 2.4 / 3.0 | 3.5941 |
| 4.5 | 165 | 9x35 | 2.75 | 3.3 / 3.0 | 3.5941 |
| 5.7 | 323 | 17x35 | 3.88 | 4.6 / 3.0 | 3.6018 |
| 7.6 | 579 | 9x35 | 5.57 | 7.0 / 3.0 | 3.5423 |

Table. 1. AWGs' parameters: $\lambda_c$ – the central wavelength, $\Delta\lambda = \lambda_{min} - \lambda_{max}$ – the device operation range, $N_{ch}$ – the number of channels, $D_{AWG}$ – the dispersion of AWG, w / h – the width and height of array waveguides, and $n_g(\lambda_c)$ – the group index of array waveguides at central wavelength.

## 4. CONCLUSION

We have presented a semi-analytical approach for estimation of phase errors and power truncation in array waveguides. This method was applied to study cumulative effect of effective index error and power truncation on crosstalk of graded-index SiGe/Si AWGs operating in mid-infrared spectral range. We conclude that the impact of effective index variation is more critical for AWGs with smaller operational wavelengths and power truncation should not exceed 5% for these types of devices. We evaluated the relative error of effective indices of 4.5 µm, 5.7 µm and 7.6 µm AWGs, that is $\approx 1.03 \times 10^{-5}$, $\approx 1.03 \times 10^{-5}$ and $\approx 1.72 \times 10^{-5}$, respectively. Their equivalent path length deviations in 10 mm length of the device are approximately 0.10 µm, 0.10 µm and 0.20 µm in 4.5 µm AWG, 5.7 µm AWG and 7.6 µm AWG, respectively, which is noticeably smaller compared to 10 mm length of the device.

## References

1. M. K. Smit and C. Van Dam, "PHASAR-based WDM-devices: Principles, design and applications," IEEE J. Sel. Top. Quant. Electron., 2, 236 (1996).
2. C. D. Lee, W. Chen, Q. Wang, Y.-J. Chen, W. T. Beard, D. Stone, R. F. Smith, R. Mincher, and I. R. Stewart, "The role of photomask resolution on the performance of arrayed-waveguide grating devices," Journal of Lightwave Technology, vol. 19, no. 11, pp. 1726–1733, Nov. 2001.
3. S. Pathak, M. Vanslembrouck, P. Dumon, D. Van Thourhout, P. Verheyen, G. Lepage, P. Absil, and W. Bogaerts, "Effect of Mask Discretization on Performance of Silicon Arrayed Waveguide Gratings," IEEE Photonics Technology Letters, vol. 26, no. 7, pp. 718–721, Apr. 2014.
4. K. Takada, H. Yamada, and Y. Inoue, "Origin of channel crosstalk in 100 GHz-spaced silica-based arrayed-waveguide grating multiplexer," Electronics Letters, vol. 31, no. 14, pp. 1176–1177, Jul. 1995.
5. H. Yamada, H. Sanjoh, M. Kohtoku, K. Takada, and K. Okamoto, "Measurement of phase and amplitude error distributions in arrayed-waveguide grating multi/demultiplexers based on dispersive waveguide," Journal of Lightwave Technology, vol. 18, no. 9, pp. 1309–1320, Sep. 2000.
6. W. Li and W. Ma, "Design and fabrication of a low crosstalk SiO/sub 2//Si array waveguide gratings," in 8th International Conference on Solid-State and Integrated Circuit Technology, 2006. ICSICT '06, 2006, pp. 498–500.
7. T. Goh, S. Suzuki, and A. Sugita, "Estimation of waveguide phase error in silica-based waveguides," Journal of Lightwave Technology, vol. 15, no. 11, pp. 2107–2113, Nov. 1997.
8. Y. Chu, X. Zheng, H. Zhang, X. Liu, and Y. Guo, "The impact of phase errors on arrayed waveguide gratings," IEEE Journal of Selected Topics in Quantum Electronics, vol. 8, no. 6, pp. 1122–1129, Nov. 2002.
9. T. Kamalakis, T. Sphicopoulos, and D. Syvridis, "An estimation of performance degradation due to fabrication errors in AWGs," Journal of Lightwave Technology, vol. 20, no. 9, pp. 1779–1787, Sep. 2002.
10. P. Barritault, M. Brun, P. Labeye, J.-M. Hartmann, F. Boulila, M. Carras, and S. Nicoletti, "Design, fabrication and characterization of an AWG at 4.5 µm," *Optics Express*, vol. 23, no. 20, p. 26168, Oct. 2015.
11. A. Koshkinbayeva, R. Orobtchouk, M. Brun, M. Carras, and P. Labeye, "Broad-band Source for Optical Gas Sensing at 5.6 — 5.9 µm," in 2015 European Conference on Lasers and Electro-Optics - European Quantum Electronics Conference, paper CH_3_5, 2015.
12. A. Koshkinbayeva, P. Barritault, S. Ortiz, S. Boutami, J. M. Hartmann, P. Brianceau, O. Lartigue, R. Orobtchouk, M. Brun, F. Boulila, and P. Labeye, "Impact of non-central input in NxM mid-IR arrayed waveguide gratings integrated on Si," IEEE Photonics Technology Letters, vol. PP, no. 99, pp. 1–1, 2016.
13. P. Labeye, A. Koshkinbayeva, M. Dupoy, P. Barritault, O. Lartigue, M. Fournier, J.-M. Fédéli, S. Boutami, S. Garcia, S. Nicoletti, L. Duraffourg, "Multiplexing photonic devices integrated on a silicon/germanium platform for the mid-infrared," in 2017 SPIE Photonics West Conference, Paper 10106-33.
14. M. Brun, P. Labeye, G. Grand, J.-M. Hartmann, F. Boulila, M. Carras, and S. Nicoletti, "Low loss SiGe graded index waveguides for mid-IR applications," Optics Express, 22, 508 (2014).